\documentclass[anon]{ceurart}

\usepackage[frozencache=true,cachedir=minted-cache]{minted} 

\setminted{breaklines=true}

\begin{document}

\copyrightyear{2023}
\copyrightclause{Copyright for this paper by its authors.
  Use permitted under Creative Commons License Attribution 4.0
  International (CC BY 4.0).}

\conference{EWAF'23: European Workshop on Algorithmic Fairness,
  June 06--08, 2023, Zurich, Switzerland}

\title{Fairness and Diversity in Information Access Systems}

\author[1]{Lorenzo Porcaro}[%
orcid=0000-0003-0218-5187,
email=lorenzo.porcaro@ec.europa.eu,
]
\address[1]{Joint Research Centre, European Commission, Italy}

\author[2]{Carlos Castillo}[%
orcid=0000-0003-4544-0416,
email=carlos.castillo@upf.edu,
]
\address[2]{Web Science and Social Computing Group, UPF, \& ICREA, Spain}

\author[3]{Emilia G\'{o}mez}[%
orcid=0000-0003-4983-3989,
email=emilia.gomez@ec.europa.eu,
]

\author[3,4]{Jo\~{a}o Vinagre}[%
orcid=0000-0001-6219-3977,
email=joao.vinagre@ec.europa.eu,
]
\address[3]{Joint Research Centre, European Commission, Spain}
\address[4]{University of Porto, Portugal}

%

    


\maketitle
\section{Introduction}
\noindent Among\footnote{This work has been adapted from Lorenzo Porcaro's PhD dissertation \citep{Porcaro2022c}.} the seven key requirements to achieve trustworthy AI proposed by the High-Level Expert Group on Artificial Intelligence (AI-HLEG) established by the European Commission (EC), the fifth requirement (``Diversity, non-discrimination and fairness'') declares: ``In order to achieve Trustworthy AI, we must enable inclusion and diversity throughout the entire AI system's life cycle. [...]  This requirement is closely linked with the principle of fairness''\citeyearpar[Chapter 2, Section 1.5, AI-HLEG,][]{AIHLEG2019}.
Hereafter, we try to shed light on how closely these two distinct concepts, \textit{diversity} and \textit{fairness}, may be treated by focusing on information access systems \citep{Ekstrand2021} and ranking literature \citep{Castillo2018, Zehlike2022, Patro2022}.
These concepts should not be used interchangeably because they do represent two different values, but what we argue is that they also cannot be considered totally unrelated or divergent. 
Having diversity does not imply fairness, but fostering diversity can effectively lead to fair outcomes, an intuition behind several methods proposed to mitigate the disparate impact of information access systems, i.e. recommender systems and search engines \citep{Celis2016, Lherisson2017, LiuBurke2018, McDonald2022}.

\section{Links between Fairness and Diversity}
The first link can be found between the concepts of \textit{group fairness} and \textit{egalitarian diversity} \citep{Drosou2017, Mitchell2020}. 
Indeed, the former, often referred to as demographic or statistical parity, is achieved when different groups, e.g., with regard to certain demographics, receive similar treatments. 
To maximise egalitarian diversity, hence having a population uniformly distributed among different groups \cite{Steel2018}, is identical to enforcing group fairness, wherein every group has equal representation i.e. similar treatment. 
This idea is behind the use of diversity constraints while intervening in the outcome of an automated decision-making system \cite{Zehlike2022}.
Moreover, group fairness relates to the concept of \textit{coverage-based diversity}, an aggregated diversity metric often used in Recommender Systems literature. 
Indeed, such metric is maximised when different groups of items are represented in the most heterogeneous way.

Second, both fairness and diversity relate to the treatment of, and consequently the impact on, protected/disadvantaged/minority groups (or classes). 
The definition of protected class is usually dependent upon laws and policies which may vary between countries, aiming at preventing any form of discrimination towards such classes. For instance, the \textit{EU Charter of Fundamental Rights} states that:
``Any discrimination based on any ground such as sex, race, colour, ethnic or social origin, genetic features, language, religion or belief, political or any other opinion, membership of a national minority, property, birth, disability, age or sexual orientation shall be prohibited'' \citeyearpar[Article 21, EC,][]{EC2012}.

As argued by Castillo \cite{Castillo2018}, ensuring fairness can be seen as ``emphasising not the presence of various groups but ensuring that those in protected groups are effectively included''. 
Under this lens, it is evident that the construction of a group diverse in egalitarian terms may not result in a fair representation if disadvantaged classes are not effectively included.
However, if we consider the exposure diversity with \textit{adversarial perspective} as defined by Helberger \cite{Helberger2018}, it explicitly aims at ``promoting exposure to critical voices and disadvantaged views that otherwise might be silenced in the public debate''. 
If defined as above, we notice that both fairness and diversity stress the importance of targeting a representation that is not only equal in terms of distribution but also that may give exposure to historically disadvantaged groups. 
We can further relate these concepts with the idea of \textit{normative diversity} \citep{Steel2018}.
Indeed, if we imagine a scenario where the non-diverse norm coincides with the privileged group --- for instance, the STEM community where the old-white-male represents the stereotype of the scientist --- increasing the diversity in a normative sense would result in a wider inclusion of marginalised voices, which is what the exposure diversity under an adversarial perspective would target.

\section{Differences and Limitations}
So far we have discussed some intersections between diversity and fairness concepts, but in order to better clarify their nature it is useful to focus also on the differences between them. 
Early quantitative definitions of both values have been proposed several decades ago, but in their rationale we note a substantial difference. 
Indeed, whilst since the beginning fairness metrics have been proposed to tackle societal issues \citep{Hutchinson2019}, most of the diversity indexes still widely used have been proposed in disparate fields, e.g., Simpson's Index in Ecology \citep{Simpson1949}, and they have been originally formulated to measure diversity intended as heterogeneity, variety or entropy, e.g., Shannon's Index \citep{Shannon1948}. 
Even if this does not undermine their use in measuring diversity, it is also true that their application needs to be contextualised for supporting the validity of the inferred results. 
Similarly, a lack of a value-oriented approach can be found in the design of the diversification techniques \citep{Carbonell1998, Smyth2001}. 
Indeed, looking at the early proposals of the Information Retrieval and Recommender Systems communities, the main goal for diversifying is to tackle the problem of ambiguity of a query or the redundancy of the results, and also to deal with uncertainty.
Great advancements have been made in this direction \cite{Castells2022}, but this utility-oriented definition of diversity has partly created ambiguity over the concept of diversity itself, at least in the communities where such approaches have been applied.

\section{Conclusion}
Whilst the aforementioned points are just a few among the several aspects that link diversity and fairness, we conclude by stressing their relevance in recent policies proposed in the European context.
The Digital Service Act (DSA) \cite{DSA}
mandates that digital services powered by technologies such as recommender systems and search engines should be monitored to guarantee the avoidance of unfair or arbitrary outcomes. 

Under a different lens, the Artificial Intelligence Act (AI Act) proposal \cite{ec2021} also refers to the need for bias monitoring as part of the mandatory requirements for high-risk AI systems. 
Moreover, in terms of diversity the AI Act explicitly states that providers of AI systems 
should be encouraged to create code of conduct 
covering aspects such as accessibility, stakeholders participation and ensuring diversity of development teams.
These two goals considered above, i.e. system-centric (ensuring bias and fairness in algorithmic systems) and a people-centric view (ensuring diversity of persons involved in the AI design process), are strongly related. 
Only fostering the diversity of development teams, and therefore embedding different perspectives, could lead to a future where Information Access Systems act in a trustworthy and fair way.

\begin{acknowledgments}
This work is partially supported by the HUMAINT programme (Human Behaviour and Machine Intelligence), Joint Research Centre, European Commission. 
The project leading to these results received funding ``la Caixa'' Foundation (ID 100010434), under agreement LCF/PR/PR16/51110009, an from 
EU-funded projects ``SoBigData++'' (grant agreement 871042) and ``FINDHR'' (grant agreement 101070212).
\end{acknowledgments}

\bibliography{main}

\end{document}